\documentclass[%
aip,
reprint,
superscriptaddress,
amsmath,amssymb,
floatfix,
]{revtex4-2}

\usepackage{graphicx}
\usepackage{dcolumn}
\usepackage{bm}

\usepackage[utf8]{inputenc}
\usepackage[T1]{fontenc}
\usepackage[english]{babel}
\usepackage{amsfonts}
\usepackage{amssymb}
\usepackage{makeidx}
\usepackage{pgf,tikz}
\usepackage{listings}
\usepackage{amsmath, amsthm}
\usepackage{placeins}
\usepackage{graphicx}
\usepackage{dsfont}
\usepackage{wrapfig}
\usetikzlibrary{arrows}
\usepackage{stmaryrd}
\usepackage{lineno}

\usepackage{url,csquotes}
\usepackage[linktocpage=true]{hyperref}
\hypersetup{colorlinks=true, citecolor=black, filecolor=black, linkcolor=black, urlcolor=blue}

\begin{document}

\title{Encoding information onto the charge and spin state of a paramagnetic atom using MgO tunneling spintronics}

\author{Mathieu Lamblin}
    \email{mathieu.lamblin@lpmmc.cnrs.fr}
    \affiliation{Institut de Physique et Chimie des Matériaux de Strasbourg, UMR 7504 CNRS, Université de Strasbourg, 23 Rue du Lœss, BP 43, 67034 Strasbourg, France}
    
\author{Victor Da Costa}
    \affiliation{Institut de Physique et Chimie des Matériaux de Strasbourg, UMR 7504 CNRS, Université de Strasbourg, 23 Rue du Lœss, BP 43, 67034 Strasbourg, France}

\author{Loic Joly}
    \affiliation{Institut de Physique et Chimie des Matériaux de Strasbourg, UMR 7504 CNRS, Université de Strasbourg, 23 Rue du Lœss, BP 43, 67034 Strasbourg, France}

\author{Bhavishya Chowrira}
    \affiliation{Institut de Physique et Chimie des Matériaux de Strasbourg, UMR 7504 CNRS, Université de Strasbourg, 23 Rue du Lœss, BP 43, 67034 Strasbourg, France}

\author{Léo Petitdemange}
    \affiliation{Institut Jean Lamour UMR 7198 CNRS Université de Lorraine BP 70239, Vandoeuvre les Nancy 54506, France}

\author{Bertrand Vileno}
    \affiliation{Institut de Chimie UMR 7177 CNRS Université de Strasbourg 4 Rue Blaise Pascal, CS 90032, Strasbourg 67081, France}

\author{Romain Bernard}
    \affiliation{Institut de Physique et Chimie des Matériaux de Strasbourg, UMR 7504 CNRS, Université de Strasbourg, 23 Rue du Lœss, BP 43, 67034 Strasbourg, France}
    
\author{Benoit Gobaut}
    \affiliation{Institut de Physique et Chimie des Matériaux de Strasbourg, UMR 7504 CNRS, Université de Strasbourg, 23 Rue du Lœss, BP 43, 67034 Strasbourg, France}
   
\author{Samy Boukari}
    \affiliation{Institut de Physique et Chimie des Matériaux de Strasbourg, UMR 7504 CNRS, Université de Strasbourg, 23 Rue du Lœss, BP 43, 67034 Strasbourg, France}

\author{Wolfgang Weber}
    \affiliation{Institut de Physique et Chimie des Matériaux de Strasbourg, UMR 7504 CNRS, Université de Strasbourg, 23 Rue du Lœss, BP 43, 67034 Strasbourg, France}
    
\author{Michel Hehn}
    \affiliation{Institut Jean Lamour UMR 7198 CNRS Université de Lorraine BP 70239, Vandoeuvre les Nancy 54506, France}

\author{Daniel Lacour}
    \affiliation{Institut Jean Lamour UMR 7198 CNRS Université de Lorraine BP 70239, Vandoeuvre les Nancy 54506, France}

\author{Martin Bowen}
    \affiliation{Institut de Physique et Chimie des Matériaux de Strasbourg, UMR 7504 CNRS, Université de Strasbourg, 23 Rue du Lœss, BP 43, 67034 Strasbourg, France}
    \email{martin.bowen@ipcms.unistra.fr}

\date{\today}

\begin{abstract}

An electrical current that flows across individual atoms can generate exotic quantum transport signatures in model junctions built using atomic tip or lateral techniques. So far, however, a viable industrial pathway for atom-driven devices has been lacking. Here, we demonstrate that a commercialized device platform can fill this nanotechnological gap. According to conducting tip atomic force microscopy, inserting C atoms into the MgO barrier of a magnetic tunnel junction generates nanotransport paths. Within magnetotransport experiments, this results in quantum interferences, and in Pauli spin blockade effects linked to tunneling magnetoresistance peaks that can be electrically controlled. We report an additional persistent memory effect that we attribute to the charging of a single "gating" C atom that is adjacent to a single C atom forming the microscale junction's effective nanotranport path. Local magnetometry experiments confirm the secondary role of magnetic stray fields on the C atoms. Our results show that, to exhibit atom-level properties, a device need not be nanoscaled, and position MgO tunneling spintronics as a promising platform to industrially implement quantum technologies.


\end{abstract}


\maketitle




Fundamental research on model atomic and molecular junctions has strongly progressed in the last two decades thanks to atomic tip and lateral junction building techniques \cite{Li2023, McCreery2022, Peng2021}. Although these technical approaches do not promote industrial nanotechnologies, they have helped reveal intriguing electronic transport mechanisms. Among them are Coulomb blockade inside single-atom transistors \cite{ryndyk_electron-electron_2016}, Coulomb drag and co-tunneling effects in capacitively coupled quantum dots \cite{keller_cotunneling_2016}, Franck-Condon blockade within carbon nanotubes \cite{Burzuri2014}, spin-phonon coupling in single-molecule magnets \cite{Ganzhorn2013}, Kondo effect with spin-oriented molecules \cite{Parks2010}, memristance and hysteresis linked to resistive switching \cite{miyamachi_robust_2012, Molen2010} and strong current fluctuations caused by vibrational coupling and structural changes \cite{secker_resonant_2011}. More recently, quantum phenomena involving internal coherence and superposition have been reported, such as phonon interference \cite{Zeng2020}, quantum interference and decoherence \cite{Greenwald2020}.

If 3$d$ metal or 2D van der Waals ferromagnetic electrodes are used to establish a fixed spin referential, then the resulting vertical or lateral spintronic device can better exploit the electron spin to examine Coulomb blockade \cite{AdvancedMaterials_2018_Mouafo_Dayen_Anisotropic,PhysicalReview_2019_Escolar_Withers_Anisotropic,AIPAdvances_2021_Suzuki_Ohya_Unconventional}, encode quantum information \cite{katcko_encoding_2021} or harvest thermal energy using discrete spin states \cite{katcko_spin-driven_2019, chowrira2020spintronic}. While molecules offer elegant means of inserting atom-level electronic states within a device, molecular spintronic technologies are still rather rudimentary and thus very far from industrial deployment \cite{barraud_phthalocyanine_2016, katcko_encoding_2021}. Finally, much research has focused on transposing robust state changes to a junction's spintronic response from more exotic barriers such as SrTiO$_3$ \cite{bowen_bias-crafted_2006} to the most widely developed industrial spintronic platform: the FeCoB/MgO/FeCoB magnetic tunnel junction (MTJ), with applications ranging from next-generation memories \cite{Kent2015} and neuromorphic computing \cite{torrejon_neuromorphic_2017} to agile microwave emitters and artificial energy harvesting \cite{zhang_ultrahigh_2018}.

\begin{figure*}[t]
    \includegraphics[width=0.99\textwidth]{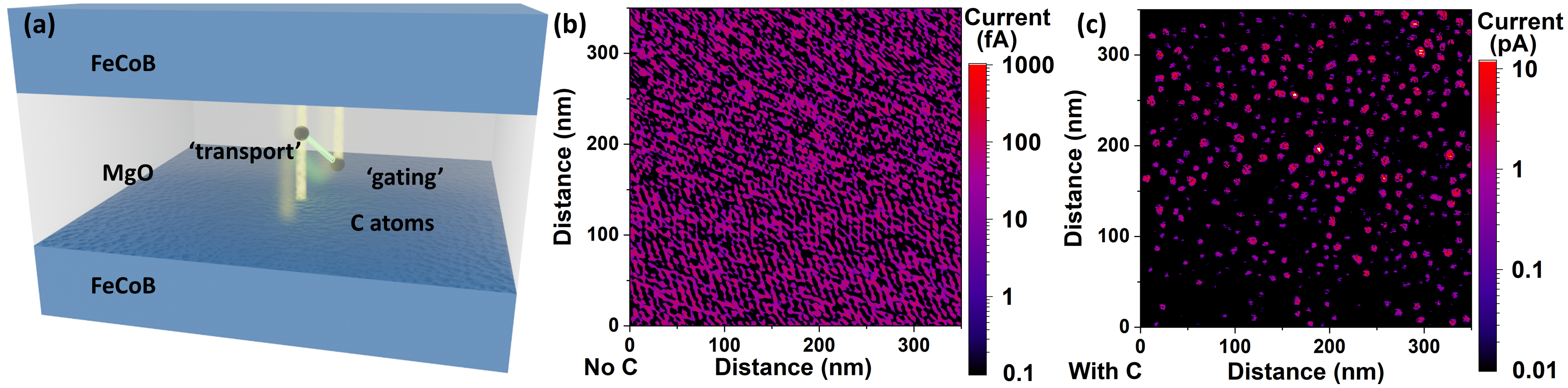}
    \caption{\textbf{Atomic nanotransport paths}. (a) Schematic of a nanotransport path formed by a 'transport' C atom that bridges the two electrodes of a microscale magnetic tunnel junction. A partly bridged path to a 'gating' C atom is also shown. Capacitive coupling and quantum interference effects are denoted by the green wiggle between the two atoms. Conducting tip Atomic Force Microscopy (AFM) scans of a half-MTJ sample (b) without C atoms in MgO and (c) with C atoms in MgO 0.9 nm away from the lower MgO interface. The interference patterns in panel (b) are due to noise at the background limit. The $\sim$10nm nanotransport lateral size in panel (c) is the result of tip convolution.}  
    \label{AFMFig}
\end{figure*}

In this paper, we demonstrate several of the aforementioned atomic transport effects using the industrialization vector that is MgO tunneling spintronics \cite{Kent2015}. Prior literature on MgO tunneling indicates that structural defects can generate localized states within the tunnel barrier \cite{Freitas2007,lu_spin-polarized_2009, Schleicher2014, Studniarek2017, schleicher_consolidated_2019,npjSpintronics_2025_Kandpal_Bowen_Oxygen}. The oxygen vacancies within the tunnel barrier generate discrete electronic states that play a role in spin transfer torque \cite{npjSpintronics_2025_Kandpal_Bowen_Oxygen}, and which we have identified \cite{schleicher_consolidated_2019} and controlled \cite{kim_control_2010,schleicher_mgo_2015,taudul_tunneling_2017} in our MgO junctions. These oxygen vacancies are diamagnetic. To achieve discrete, unpaired electron spin states, we introduce carbon atoms onto these oxygen vacancies \cite{katcko_spin-driven_2019} (see Methods). These states, and the nanotransport paths they promote across the MgO MTJ, are schematized in Fig.~\ref{AFMFig}(a).

\section{C-borne nanotransport paths across a MgO tunnel barrier}

We first use conducting Atomic Force Microscopy (AFM) to examine how inserting C alters the nanoscale conductivity across the MgO barrier (see Methods). When C is absent (Fig.~\ref{AFMFig}(b)), no measurable current is detected by the tip. Assuming Gaussian noise, the standard deviation of the measured current is $\sigma\approx0.065$ pA. Upon inserting C into MgO, we observe a series of current hotspots up to 120 pA that statistically contribute to the tail of the current distribution \cite{DaCosta2000}. The $\sim$10 nm apparent lateral size is due to tip convolution over the several-atom-wide effective zone \cite{taudul_tunneling_2017,katcko_spin-driven_2019} of the nanotransport path \cite{Studniarek2017}. Since quantum mechanical tunneling scales exponentially with the barrier's spatial width and energy height, the most prominent hotspot will funnel all the current across a microscale tunnel junction \cite{Studniarek2017}, as schematized in Fig.~\ref{AFMFig}(a). It is this mechanism that enables atom-level properties in microelectronics here.

\begin{figure*}
    \includegraphics[width=0.85\textwidth]{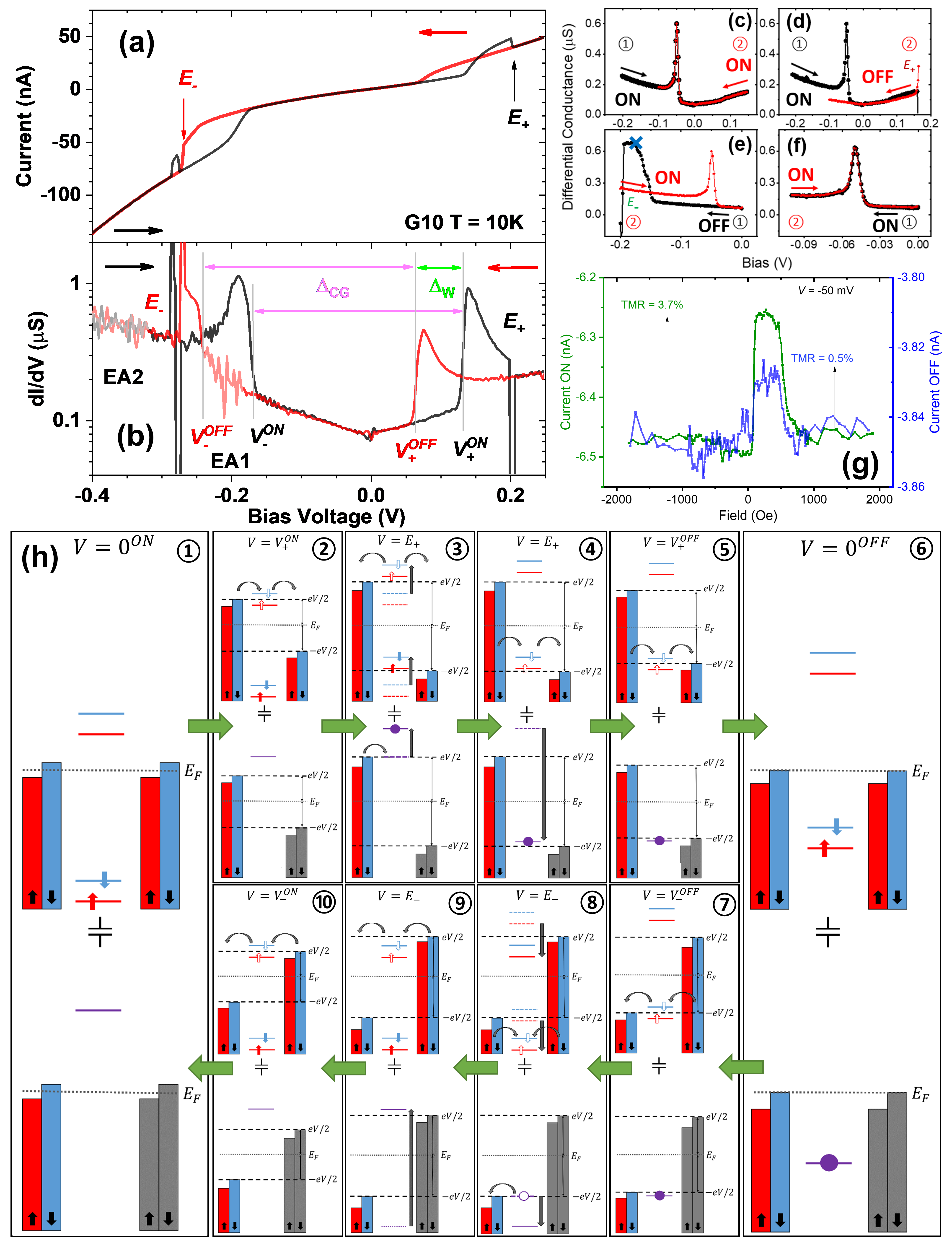}
    \caption{\textbf{Memristive Coulomb blockade at the atomic level.} (a) $IV$ and (b) differential conductance d$I$/d$V$ data at $T = 10$ K on junction G10. The $E_+$ and $E_-$ writing events cause a shift $\Delta_W = 73$ mV in the otherwise constant energy gap $\Delta_{CG} = 310$ mV between conductance peaks. Transport noise due to interference with an environmental 'gating' atom are shown using semi-transparent datapoints. (c-f) d$I$/d$V$ data upon sweeping bias to test the interplay between the presence of the Coulomb blockade peak and the writing events $E_-$ and $E_+$. Starting in the black-coded junction state, bias is swept (black data of branch 1), and then reversed (red data of branch 2). Reaching a writing event while in the black state impacts the presence of the conductance peak. The blue cross in panel (e) pinpoints the bias and junction state in which the data of Fig.~\ref{noise}(a) was acquired. (g) Magnetoresistance traces at $V= - 50$ mV in the ON/OFF junction states. (h) Model of a memristive Coulomb blockade involving at least one atom in the transport path, and at least one environmental 'gating' atom that is capacitively linked to the transport atom but doesn't participate directly in electronic transport across the junction (see also Fig~\ref{AFMFig}(a)). The various operations of the memristive cycle, showing how charging/discharging the gating atom controls the energy levels of the transport atom, are shown.}
    \label{mem1}
\end{figure*}



\section{Atom-level information encoding in a microscale device}

Within a MgO MTJ (see Methods), the C-borne spin states can be electrically manipulated to encode information. The cyclical $IV$ traces in Fig.~\ref{mem1}(a) show two ON and OFF current branches that are linked through 'writing' events $E_-$ and $E_+$. Below these writing thresholds, the differential conductance (d$I$/d$V$; panel (b)) exhibits a set of two peaks whose bias positions rigidly shift by $\Delta_W \approx$ 70 mV between the ON and OFF branches. Carefully crafted bias sweeps (panels (c-f)) below or above a writing threshold reveal how the writing event causes the conductance peak presence to change.  As discussed in SI Note 4, the constant voltage gap $\Delta_{CG} = 310$ mV between the pair of conductance peaks can be tracked across several other junction states. Note how the current increases exponentially within this voltage gap. 

These features are absent in MgO MTJs without C, \textit{i.e.} when the nanotransport path only involves oxygen vacancies \cite{taudul_tunneling_2017,Schleicher2014,schleicher_consolidated_2019}. Also, while O$^{2-}$ electromigration across transition metal oxides can explain resistive switching \cite{hwang_resistive_2010}, the ionic nature of MgO and low electric fields here preclude \cite{bertin_random_2011,aguiar-hualde_taming_2014} this explanation for the writing events. We therefore attribute these conductance peaks to the edges of the Coulomb blockade regime across the MgO MTJ thanks to electronic states of C atoms in the MgO barrier. Then, the writing-induced energy shift of the Coulomb gap may be due to charging environmental 'gating' C atoms that do not participate in transport, but are capacitively coupled to the 'transport' C atoms (see Fig.~\ref{AFMFig}(a)) and essentially act as an electrostatic gate. This is supported in Fig.~\ref{mem1}(a) by the presence of noisy regions EA1 and EA2 (semi-transparent datapoints). Indeed, the spectral position and width of these regions do not correspond to Coulomb blockade peaks in the present dataset, but rather in other datasets acquired on the same junction but with different 'transport' and 'gating' attributions of the C atoms upon thermal cycling. See SI Note 4 for details. 

Within an appropriate capacitance model (see SI Note 5), the Coulomb gap $\Delta_{CG} = 310$ mV indicates a 0.2 nm radius for the transport quantum dot: transport is proceeding across the electronic states of an individual carbon atom.  Furthermore, if the voltage shift $\Delta_W \approx$ 70 mV between junction states corresponds to changing the 'gating' atom's charge by 1 electron, then we infer that it is positioned at twice the radius away from the 'transport' C atom (see SI Note 5).

A similar hysteretic behavior induced by conductance jumps has been reported in model junctions \cite{jia_covalently_2016, Lortscher2006, schwarz_field-induced_2016,wu_conductance_2008, leoni_controlling_2011, gerhard_electrically_2017, zhang_bi-stable_2020}, 
including abrupt switching between two conductance branches due to electron charging \cite{wu_conductance_2008}. 
Inspired by this literature, we present in Fig.~\ref{mem1}(h) the schematic of a two-quantum dot model that can explain this memristive Coulomb blockade behavior in our microjunction. The upper 'transport' quantum dot (TQD) is connected to both leads, while the lower 'gating' quantum dot (GQD) is connected only to the left lead. Both quantum dots are capacitively coupled together. 

Starting at $V = 0$ in the ON branch, the lower-lying levels of the TDQ are filled, the upper-lying are empty, and the single level of the GDQ is empty (see Fig.~\ref{mem1}(h)1). At $V = V_+^{ON}$ (panel 2), sequential tunneling from the left lead onto the upper level of the TDQ occurs, leading to the conductance peak. 
Increasing the voltage to $V = E_+$ causes potential alignment with the GDQ level (panel 3). Electronic charging from the left electrode makes the levels of both QDs rise (step 1): the TQD's higher levels are above the potential, while the lower levels lie between the potential window offered by the electrodes. After equilibration, electron escape from the TQD occurs, and sequential tunneling starts through the lower levels. Within a narrow energy window, either the system trivially returns from 3 to 2, or the new dynamic filling of the TQD's energy levels capacitively lowers the GDQ's level (see Fig.~\ref{mem1}(h)4). This capacitive back-action traps the electron that filled the GDQ level, preventing it from jumping back onto the electrode. This causes a conductance jump from the ON to OFF branch, and only a negative potential can release this trapped electron and revert the junction back to the ON state.


If the voltage is now decreased to $V_+^{OFF}$ (see Fig.~\ref{mem1}(h)5), another conductance peak is observed due to sequential tunneling through the lower lying level of the TQD. Then, at $V=0$ (see Fig.~\ref{mem1}(h)6), the Coulomb blockade region is recovered but in an energy window that is shifted by the presence of the electron on the GDQ.

Similar arguments can describe the process of electron escape from the GDQ and the switch from the OFF back to the ON transport regime (see panels 7-10 of Fig.~\ref{mem1}(h)). This two-QD capacitive charging cycle, which describes the experimental data of Fig.~\ref{mem1}(b), can also be adjusted for other cases. The examples of Fig.~\ref{mem1}(d) and SI Fig.~S4 are fulfilled if the upper levels of the TQD lie below the empty level of the GDQ in the ON state. Then, an electron tunneling from the left electrode would address the GDQ level around $V = E_-$ before reaching the edge of the Coulomb blockade region.

\section{Quantum Interference effects}

\begin{figure}[t]
    \includegraphics[width=0.48\textwidth]{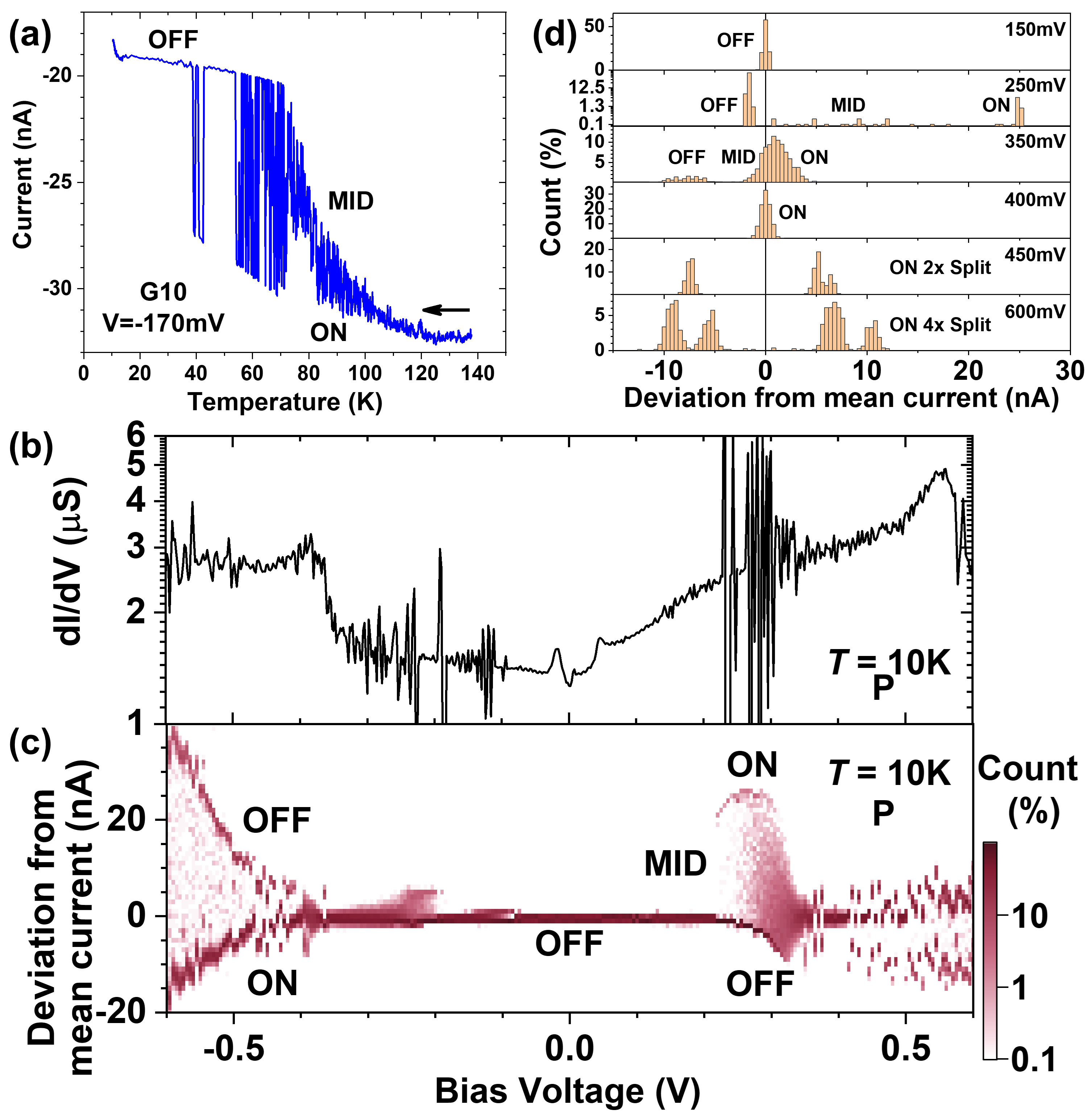}
    \caption{\textbf{Quantum transport interference between individual carbon atoms.} (a) Intensity as function of temperature at -2000 Oe and -170 mV for junction G10. (b-d): junction C5. (b) Differential conductance at 10 K for the P magnetic state. (c) Color map of the statistic weight of the current deviation from the mean current as a function of applied bias. The junction P magnetic state was used. At each bias value, 1000 current measurements were acquired. Datasets at select voltage values are shown in panel (d).}  
    \label{noise}
\end{figure}


The discrete energy levels of the transport and gating quantum dots give rise to quantum transport interference effects. We present in Fig.~\ref{noise}(a) the temperature dependence of the current measured on junction G10 at $V = -170$ mV. As emphasized by the blue cross in Fig.~\ref{mem1}(e), the junction is bistable at 10 K at this bias. Upon cooling down from 140 K, the current broadens into two ON/OFF branches, with intermediate spectral weight of low intensity (MID branch). Further evidence of these regimes appears in SI Note 7. Since temperature alone can promote a dominant transport branch, we conclude that electron-phonon interactions, \textit{i.e.} vibrons, are also involved in quantum interference transport here, in line with prior literature \cite{koch_theory_2006, galperin_molecular_2007, mozyrsky_intermittent_2006, aradhya_single-molecule_2013}.

We use bias-dependent current statistics to observe transport branches due to quantum interference. We first plot in Fig.~\ref{noise}(b) differential conductance data 
of junction C5 at $T = 10$ K. On the log scale, we observe sizable increases in conductance, and the clear appearance of noise for $V < -0.4$ V and for $V > 0.3$ V. Within $-0.6 < V$ (V) $< +0.6$, after setting the dc applied bias, we measured the current 1000 
times. The statistics of a given current deviation from the mean current as a function of applied bias are plotted in panel (c). Within $|V| < 0.2$ V, the current remains essentially stable along the OFF branch. For  $|V| > 0.2$ V, the transition to the MID state can appear. This is especially visible within $0.2 < V$ (V) $< 0.4$. While the OFF branch shifts to values below the mean, the MID branch collapses into the dominant branch around $V = 0.4$ V. A higher resolution dataset is shown in SI Note 1. For $V > 0.4$ V, this branch splits into two ON/OFF branches. A similar branching is seen for $V < -0.4$ V. Erratically, each of these two branches can split into two subbranches, for a total of four branches (see Fig.~\ref{noise}(d)). The conductance noise (panel c) thus arises from transport interference between these branches 
which are visualized by our current statistics.


Overall, these experimental findings support a picture of vibron-mediated co-tunneling and/or Coulomb drag between nanotransport paths \cite{keller_cotunneling_2016} due to blockade effects such as Franck-Condon blockade \cite{koch_theory_2006}. According to the theory of out-of-equilibrium polaron dynamics \cite{mozyrsky_intermittent_2006, galperin_hysteresis_2005}, transport across an impurity with discrete energy levels, such as a quantum dot or a molecular orbital that is strongly coupled to a vibrational degree of freedom and that is tunnel coupled to two leads, can explain the observed multi-stability of different states and the telegraphic switching between the different intensity branches. The bias voltage acts as an effective temperature and a force that modulates the oscillator potential. Around specific voltage biases, the effective vibron energy potential landscape develops several minima, and thus promotes metastability between different occupation states. Two scenarios are possible. 1) Thermally activated switching between different intensity branches related to each of the well-separated potential minima can occur \cite{Molen2010}. 2) Large fluctuations around a single intensity branch occur when at least two minima are sufficiently degenerate to energetically trigger an interference effect \cite{wierzbicki_influence_2011, wu_control_2004}. In this latter case, intermediate states outside the well minima are measured. 

\section{Spintronic signatures}

The memristive behavior of the Coulomb blockade peak due to environmental charging represents one level of information encoding, with a write success rate $>80\%$ (see SI Note 3). To examine how the spin polarization of the charge current impacts the atom-level memristive Coulomb blockade effect and quantum interference effects, we compare transport properties upon switching the orientation of the electrodes' magnetization from parallel (P) to antiparallel (AP). Note that C atoms in MgO are paramagnetic, \textit{i.e.} can bear an unpaired spin \cite{katcko_spin-driven_2019}. 

\begin{figure}[t]
    \includegraphics[width=0.48\textwidth]{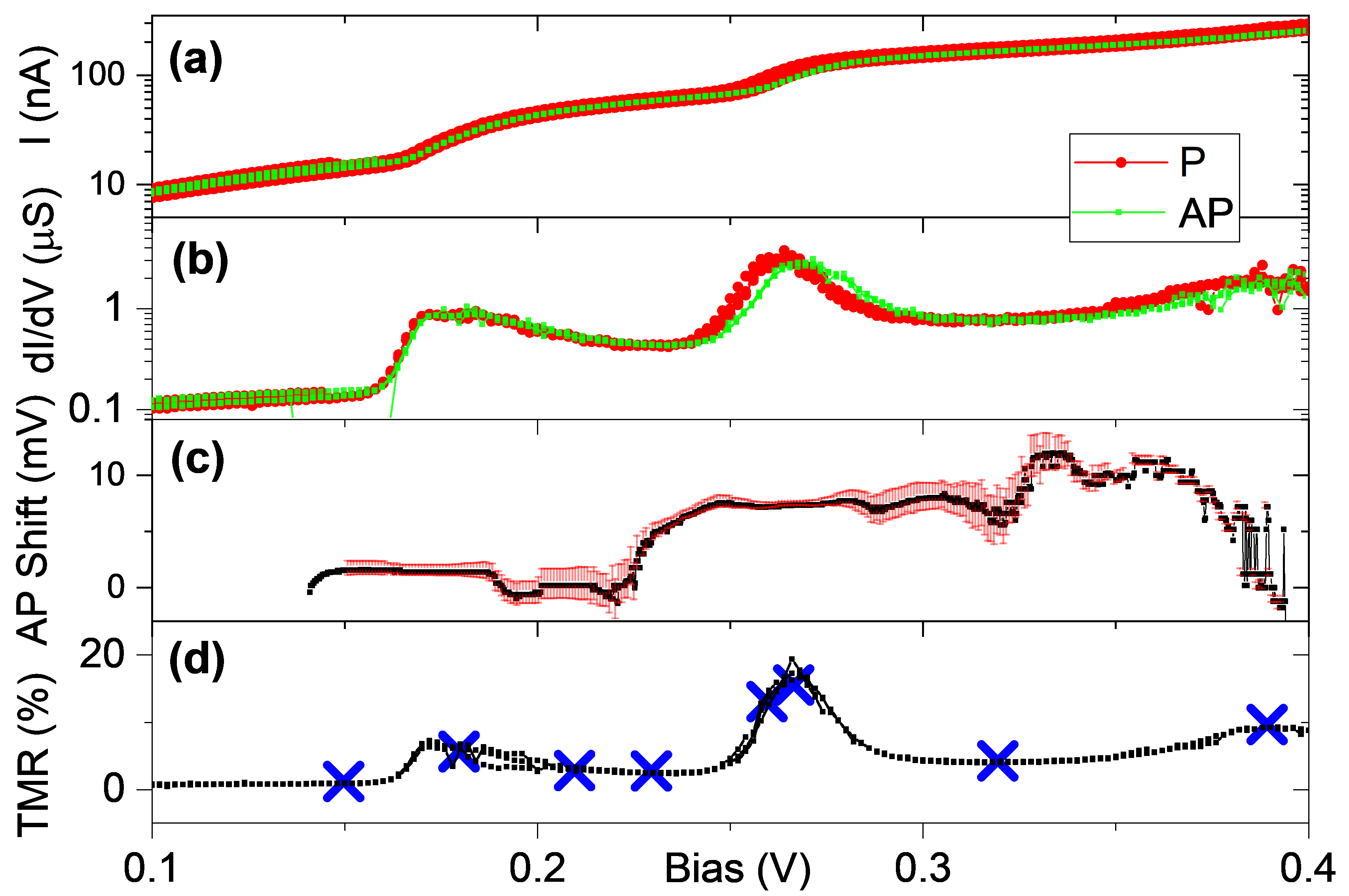}
    \caption{\textbf{Spin accumulation due to Pauli spin blockade.} (a) Current-voltage characteristic for the P and AP junction states, and (b) the resulting d$I$/d$V$ differential conductance. (c) Voltage shift between the P and AP conductance data (see Methods for details). (d) Bias dependence of the tunnel magnetoresistance (TMR) calculated from (a). Blue crosses indicate TMR amplitudes obtained from $I(H)$ experiments. All data: junction G10.
    }  
    \label{mag}
\end{figure}

We plot in Fig.~\ref{mag}(a) the $IV$ in the MTJ's P and AP states at $T = 10$ K. Both exhibit a series of plateaus and increases that are absent in junctions without C atoms \cite{Schleicher2014, Studniarek2017, schleicher_consolidated_2019}. The corresponding differential conductance (d$I$/d$V$) data (see panel (b)) reveals a series of peaks. While the P and AP datasets share similar traits, the AP plot appears to shift to higher bias as the onset of each conductance increase is reached. This is confirmed through an analysis of the correlated shift between the two datasets in Fig.~\ref{mag}(c) (see Methods).


The tunnel magnetoresistance TMR $=\frac{I_P}{I_{AP}-1}$ characterizes to what extent the two spin channels of current flowing across the device are asymmetric. In our MgO MTJs without C \cite{Schleicher2014}, we observe TMR $\approx 200$ \% at 10 K around $V = 0$, and a mostly monotonous TMR decreasing with increasing $|V|$.  When C atoms are introduced into MgO, we observe only a few \% TMR around $V = 0$. Instead, due to the shift in AP conductance to higher bias with each succeeding conductance peak, the TMR bias dependence closely mimics the junction conductance (compare panels (b) and (d) of Fig.~\ref{mag}). This unusual, highly structured TMR bias dependence \cite{AIPAdvances_2021_Suzuki_Ohya_Unconventional}
is confirmed through discrete $I(H)$ datasets at fixed $V$ (crosses in Fig.~\ref{mag}(d)). We observe local TMR maxima precisely on the conductance peaks, with an absolute maximum of 15\%.  

We present $I(H)$ data in Fig.~\ref{mem1}(g) in both the ON and OFF states of the junction, taken at $V = -50$ mV; \textit{i.e.} precisely at the bias value corresponding to the memristive Coulomb blockade peak shown in panels (c-e) of Fig.~\ref{mem1}. 
We observe that I$_P$ increases and that I$_P$-I$_{AP}$ is multiplied by 7, so that the TMR increases from 0.5\% in the OFF state to 3.7\% in the ON state due to the Coulomb blockade peak. Within the paradigm of MgO tunneling spintronics \cite{moodera_frontiers_2010}, this suggests that overcoming the Coulomb blockade effectively opens an additional transport channel for the dominant majority spin carriers.


According to Płomińska and Weymann \cite{plominska_magnetoresistive_2019}, the similar bias-dependent conductance peaks in the P and AP datasets, which yield a similar TMR trace, is the signature of a special kind of Coulomb blockade involving the Pauli exclusion principle called Pauli Spin blockade \cite{fransson_pauli_2006, kuo_theory_2011, plominska_pauli_2016}. As the electronic level of the TQD is reached with applied bias, sequential tunneling is suppressed while co-tunneling mechanisms become dominant and modify the TQD's spin state (and eventually its charge \cite{katcko_spin-driven_2019}) as a form of spintronic anisotropy \cite{misiorny_spin_2009, misiorny_transverse_2014, saygun_voltage-induced_2016}. With increasing voltage, the FeCoB/MgO tunneling-induced accumulation of mostly spin up carriers lifts the spin state degeneracy. 
This spin splitting differs between the MTJ's P and AP states due to different energy alignments of the C electronic levels \cite{plominska_magnetoresistive_2019,katcko_spin-driven_2019}. This effect is also modeled by our schematic in Fig~\ref{mem1}(h), which features this spin splitting. In the P state, the potential of the electrodes will align with the spin-split energy levels of the TQD at lower absolute bias than in the AP state. This would result in a voltage shift between the conductance peaks. SI Notes 1 and 4 provide additional evidence of this effect. 


The conductance shift of Fig.~\ref{mag}(c) is thus an experimental manifestation of spin accumulation. Spin accumulation explains the shift in quantum interference data and the spin polarization of the transport branch therein (see SI Notes 1-2). It also explains why the TMR bias dependence tracks that of junction conductance (see Fig.~\ref{mag}(d)), as well as the huge increase in I$_P$-I$_{AP}$ due to the memristive Coulomb peak (see Fig.~\ref{mem1}(g)) leading to the seven-fold TMR enhancement. The best qualitative agreement between the results of Płomińska and Weymann \cite{plominska_magnetoresistive_2019} and our experimental datasets is that of 'parallel' or 'T-shaped' electronic transport across spin states, rather than the other series scenario proposed. This supports the atomic description of the effective nanotransport path across our microscale MTJs. 



\begin{figure}[b]
    \includegraphics[width=0.48\textwidth]{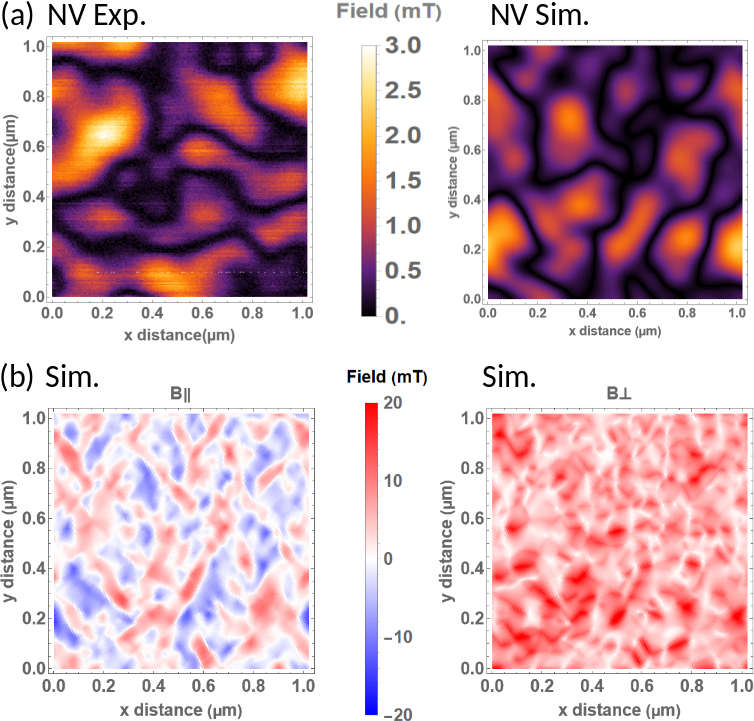}
    \caption{\textbf{Magnetic stray field on the C atom due to the ferromagnetic electrode}. (a) NV center magnetometry scan 90 nm away from the half-MTJ sample surface. The color encodes the stray field absolute value projected onto the NV center axis (100-oriented diamond tip). (b) Extrapolated maps of the stray field longitudinal $B_{||}$ and transverse $B_{\perp}$ components.} 
    \label{nv}
\end{figure}

To obtain more insight into the origin of the magnetic field experienced by the transport C atom, we performed nitrogen vacancy (NV) center microscopy experiments on a half-MTJ sample (see Methods), and simulated these data using Mumax3 \cite{Lel2014, Vansteenkiste2014} to extract micromagnetic parameters. An example of experimental data and simulation is shown in Fig.~\ref{nv}(a) for a NV center height of 90 nm. Using the same parameters, we are able to fit other NV scans at heights ranging from 60 nm to 180 nm. This allows us to confidently extrapolate the stray field at 2.5 nm above the ferromagnetic surface. Figs.~\ref{nv}(b) (left/right) show the extrapolated maps of the longitudinal $B_{||}$ / transverse $B_{\perp}$ components of the stray field \cite{paperNV}. We find that no field component exceeds 20 mT, \textit{i.e.} orders of magnitude below values inferred by the spintronic energy shifts observed in magnetotransport (see Fig.~\ref{mag}). We thus infer that the effect arises from the electronic magnetic field $\sim$50 T due to spin-polarized charge transfer from the ferromagnetic metal, in combination with bias-induced spintronic anisotropy\cite{katcko_spin-driven_2019}.

\section{Conclusion}

To conclude, inserting C atoms into ultrathin MgO layers generates localized paramagnetic \cite{katcko_spin-driven_2019} states. 
Experiments on micronic FeCoB/MgO/FeCoB magnetic tunnel junctions, which are crafted using straightforward technological processes, show that the effective nanotransport path \cite{Studniarek2017} involves individual C atoms. Their discrete energy levels promote Coulomb blockade effects that can be reproducibly shifted in energy by charging events on a neighboring C atom. The tunnel coupling between these transport and environmental 'gating' carbon atoms promotes quantum interference effects. Spin-polarized transport induces spin accumulation that further lifts the spin degeneracy of the unpaired C electron in MgO. This leads to a voltage shift in Coulomb peaks between the MTJ's P and AP datasets, and to quantum interference effects. Spin accumulation also accounts for the huge enhancement of the spintronic performance when a Coulomb peak is memristively controlled.

We thus demonstrate how to use both the electron charge and spin to encode information on an individual paramagnetic atom in a solid-state, industrializable device. These results showcase MgO tunneling spintronics as a promising industrial platform for quantum technologies at potentially practical temperatures, to deploy quantum transport effects that are normally only seen in model junctions. Naturally, compared to research on model junctions, the main difficulty here will be to engineer the C-borne paths embedded into the MTJ pillar so as to obtain 1) a higher device-to-device reproducibility at 2) with electronic stability at higher temperatures. Possible approaches include engineering the impedances that electronically link the transport and gating atoms to the magnetic tunnel junction electrodes; engineering the paramagnetic center's chemical bonds to MgO; and controlling the density of oxygen vacancies \cite{taudul_tunneling_2017}. Final applications of quantum spintronics not only encompass information encoding \cite{katcko_encoding_2021}, but also energy harvesting \cite{katcko_spin-driven_2019,chowrira2020spintronic,npjQuantumInformation_2023_Bowen_Bowen_Atomlevel} vectors.

\section{Methods}
\subsection{Stack growth and MTJ processing}

Glass//Ta(5)/Co(10)/IrMn(7.5)/CoFeB(4)\\/MgO(0.9)/C(0.6)MgO(1.7)/CoFeB(3)/Ta(2)/Pt(1) samples (all thicknesses in nm) were sputter-grown on Corning 1737 glass substrates \cite{bernos_impact_2010}. Stacks were post-annealed in an in-plane magnetic field of 200 Oe for one hour at a temperature $T_a$ of 200 °C to magnetically pin the lower electrode thanks to the IrMn antiferromagnetic layer. Samples were then processed by optical lithography \cite{halley_electrical_2008} into 20 $\mu$m-diameter MTJs, and measured on a variable-temperature magnetotransport bench. 54 junctions were tested on these samples, 35 of them were either metallic of open circuit. Among the other 19 interesting junctions, all the reported effects have been observed in several of them. 8 of them presented memristive properties with branch jumps and conductance peaks, among which at least 3 presented magnetomemristive properties correlating with the conductance, while 7 of them presented jumps and noisy behaviors typical of interferences between several nanotransport paths (see SI Note 6). 

\subsection{Conducting tip atomic force microscopy} 

Local probe experiments were carried out on a Si//Ta(3)/Pt(5)/IrMn(8)/CoFeB(5)/MgO(0.4)/C(0.6)\\/MgO(0.6) stack using an Icon AFM (Atomic Force Microscope) from Bruker. Topography images were acquired in standard contact mode (force constant) and, simultaneously, using an additional TUNA module (Bruker), which allowed current collection via a conductive tip. A voltage bias of 10 mV was applied to conductive AFM tips (Bruker SCM-PIT-V2; PtIr-coated Si with a 3N/m elastic constant) to enable electron collection. The current amplifier gain was set to 10 pA/V.

\subsection{NV center scanning magnetometry}

We conducted NV scanning magnetometry using a Qzabre microscope, which integrates scanned probe microscopy and optically detected magnetic resonance. This enabled us to image the magnetic field distribution induced by the magnetic texture of the ferromagnetic bottom electrode. The electrode stack configuration is as follows: Glass//Ta(5)/Co(10)/IrMn(7.5)/CoFeB(4)/MgO(2.5)\\/Pt(2.5). For this study, we employed a commercial Qzabre tip (QST-100). Scans were carried out at four different heights: 180 nm, 130 nm, 90 nm, and 60 nm. Below 60 nm, the image quality deteriorated, likely due to a fluorescence quenching effect. To simulate the stray field, we used the mumax3 code \cite{Lel2014, Vansteenkiste2014}, incorporating a random spatial distribution of anisotropy direction with 50 nm magnetic grains. Our mumax3 problem will be made available upon request. Further details will be provided in a more technical paper.

\subsection{Voltage shift calculation}

The voltage shift between the P and AP conductance data of Fig.~\ref{mag}(c) was calculated by doing a curve fitting of AP data over P data using a rolling window of 40mV width, with the voltage shift as the only parameter. The fit and estimation of uncertainties were done using the LMFIT python module\cite{lmfit} (Non-Linear Least-Squares Minimization and Curve-Fitting). The calculated voltage shift was then defined as the voltage shift of the mean bias value within the rolling window.

\section{Acknowledgements}
We gratefully acknowledge PhD funding for M.L. from Ecole Polytechnique. We acknowledge financial support from the ANR (ANR-21-CE50-0039), the Contrat de Plan Etat-Region grants in 2006 and 2008, by “NanoTérahertz”, a project co-funded by the ERDF 2014–2020 in Alsace (European Union fund) and by the Region Grand Est through its FRCR call, by the impact project LUE-N4S part of the French PIA project “Lorraine Université d’Excellence”, reference ANR-15IDEX-04-LUE and by the FEDER-FSE “Lorraine et Massif Vosges 2014–2020”, a European Union Program. This work of the Interdisciplinary Thematic Institute QMat, as part of the ITI 2021-2028 program of the University of Strasbourg, CNRS and Inserm, was supported by IdEx Unistra (ANR 10 IDEX 0002), and by SFRI STRAT’US project (ANR 20 SFRI 0012) and EUR QMAT ANR-17-EURE-0024 under the framework of the French Investments for the Future Program. This work was supported by France 2030 government investment plan under grant reference PEPR SPIN–MAT ANR-22-EXSP-0007 and PEPR SPIN –ADAGE ANR-22-EXSP-0006.

\bibliography{ref}


\begin{figure}[h]\vspace*{-2cm}\makebox[\textwidth]{\includegraphics[scale=1,page=1]{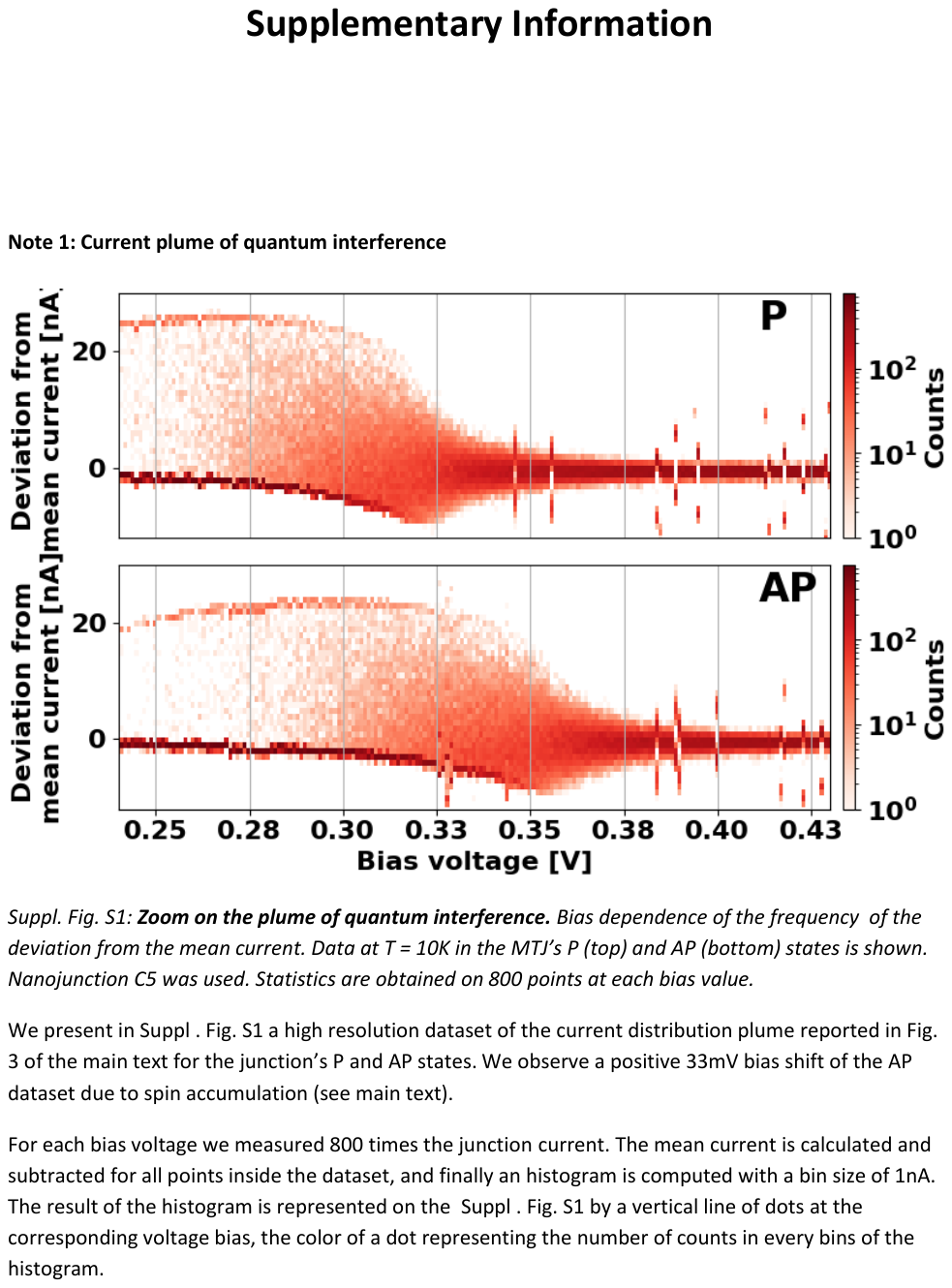}}\end{figure}

\begin{figure}[h]\vspace*{-2cm}\makebox[\textwidth]{\includegraphics[scale=1,page=2]{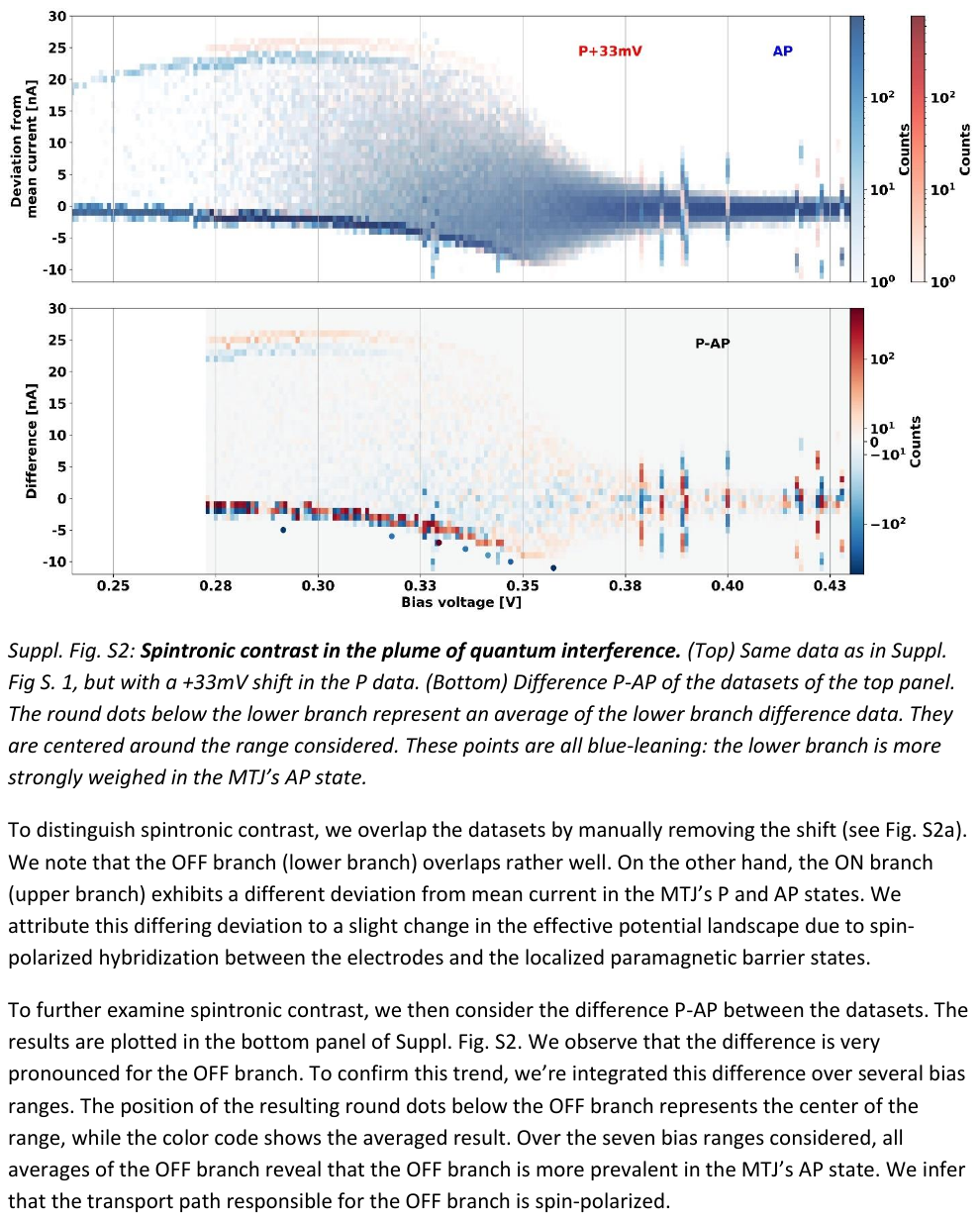}}\end{figure}

\begin{figure}[h]\vspace*{-2cm}\makebox[\textwidth]{\includegraphics[scale=1,page=3]{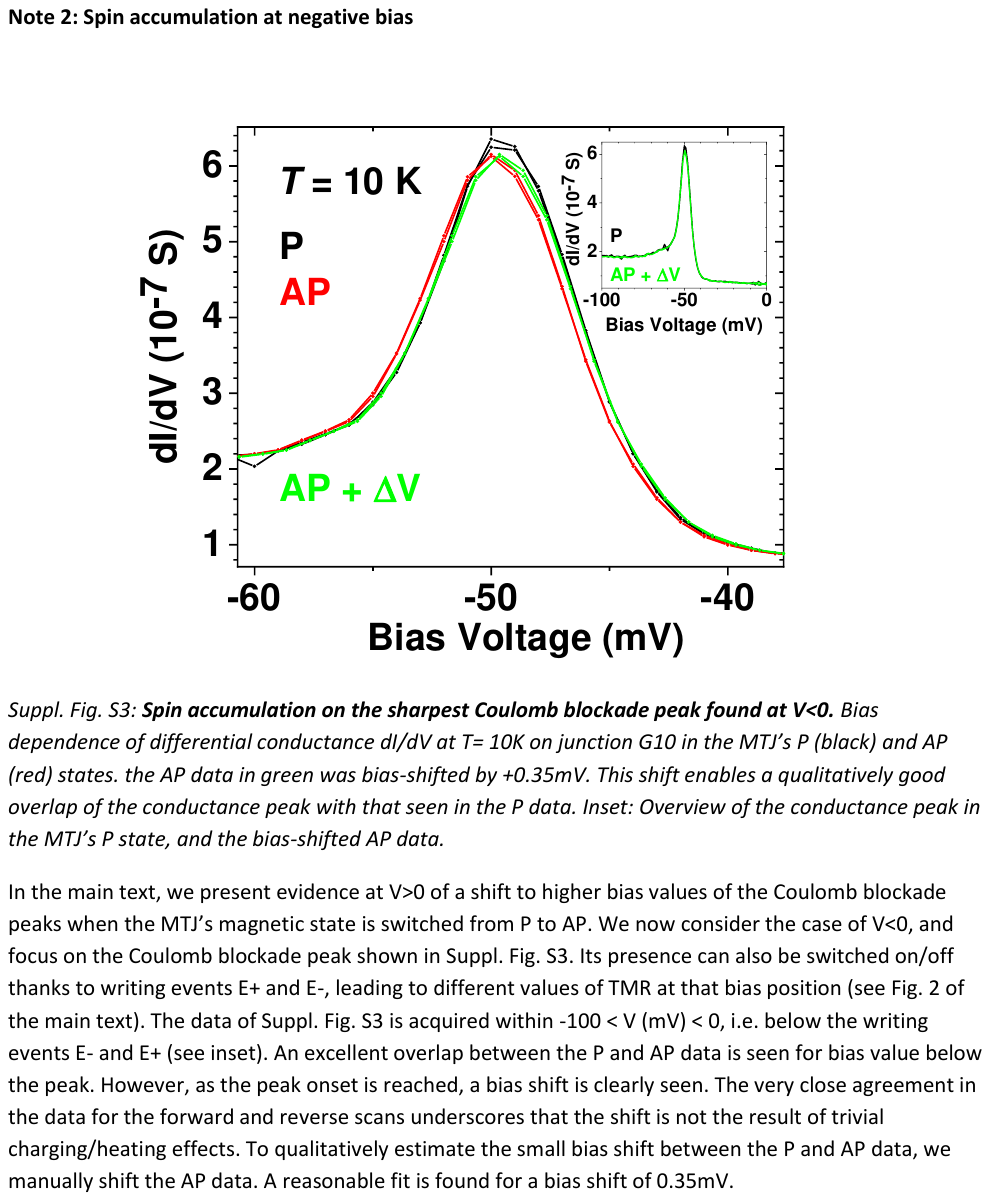}}\end{figure}

\begin{figure}[h]\vspace*{-2cm}\makebox[\textwidth]{\includegraphics[scale=1,page=4]{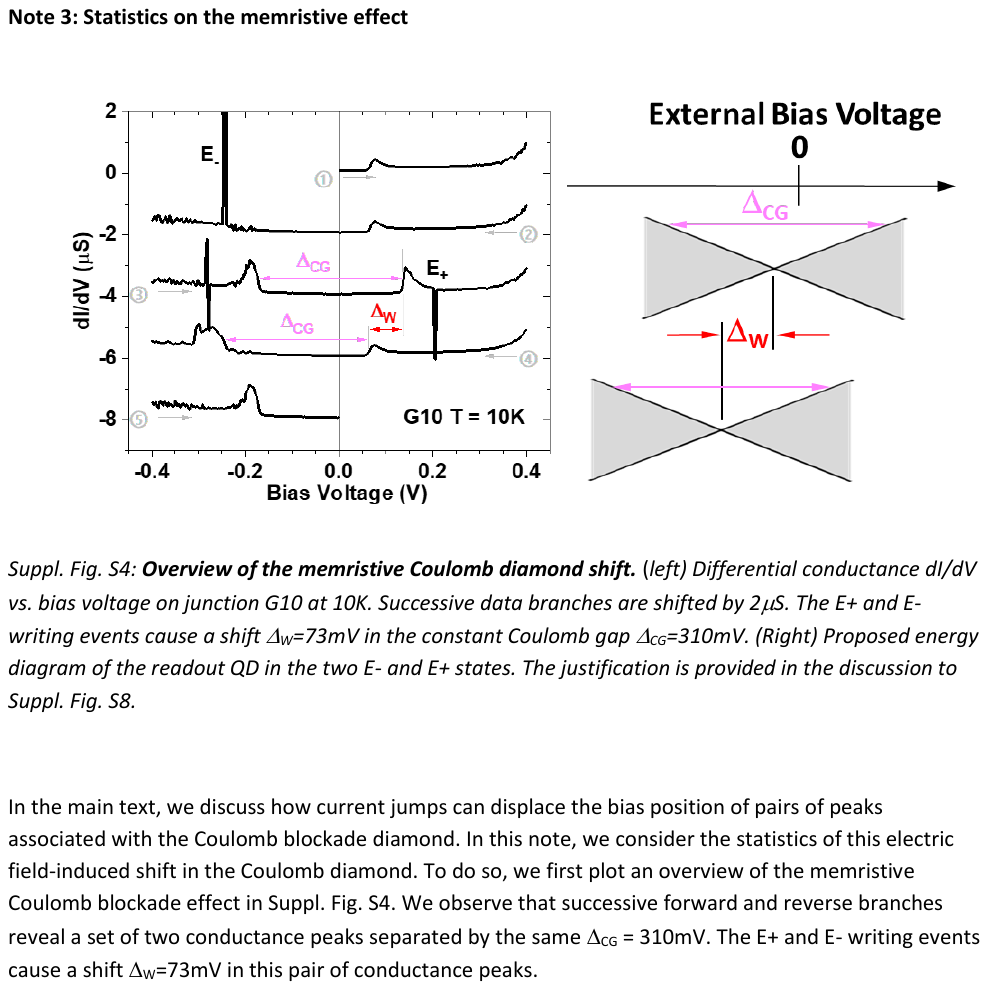}}\end{figure}

\begin{figure}[h]\vspace*{-2cm}\makebox[\textwidth]{\includegraphics[scale=1,page=5]{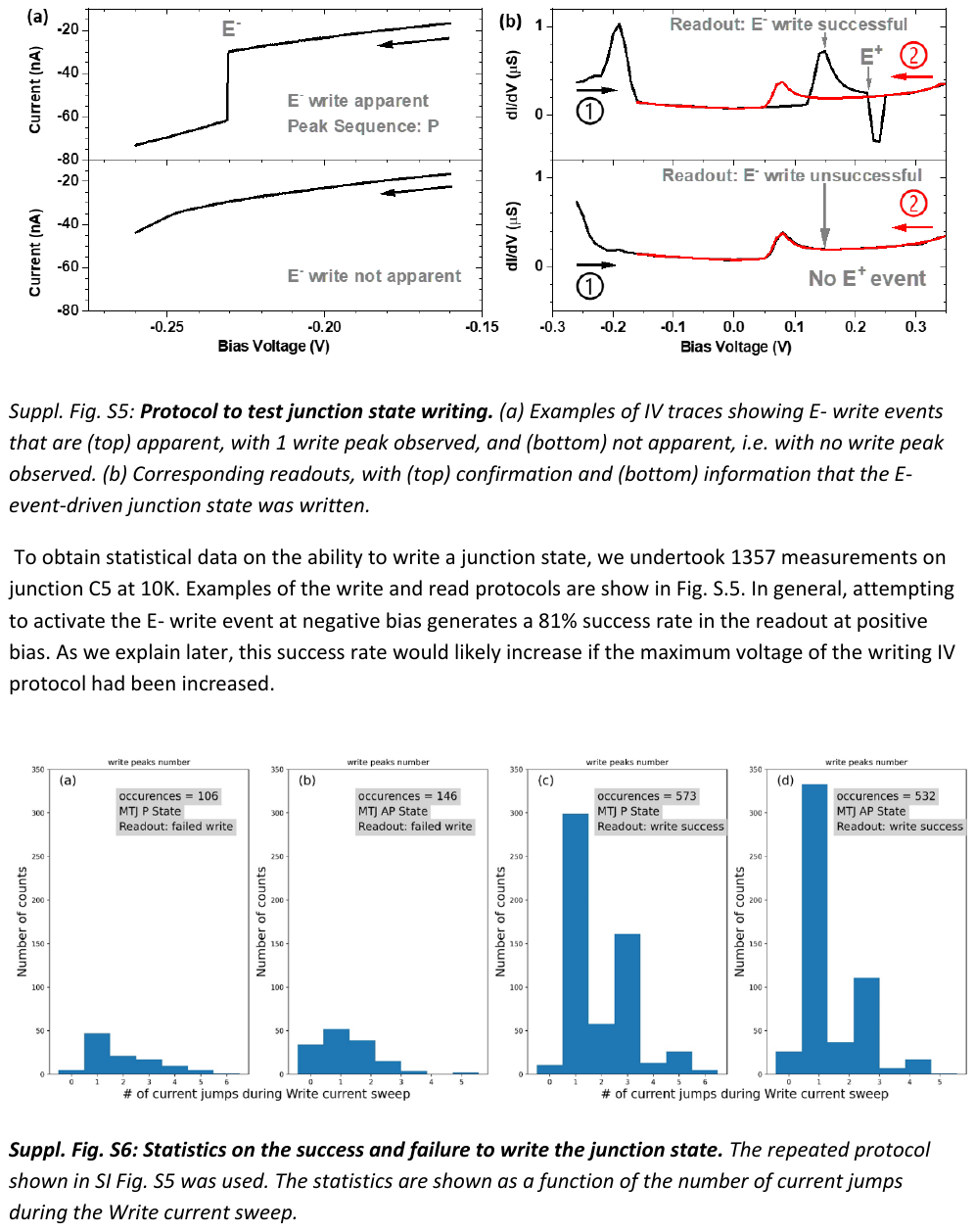}}\end{figure}

\begin{figure}[h]\vspace*{-2cm}\makebox[\textwidth]{\includegraphics[scale=1,page=6]{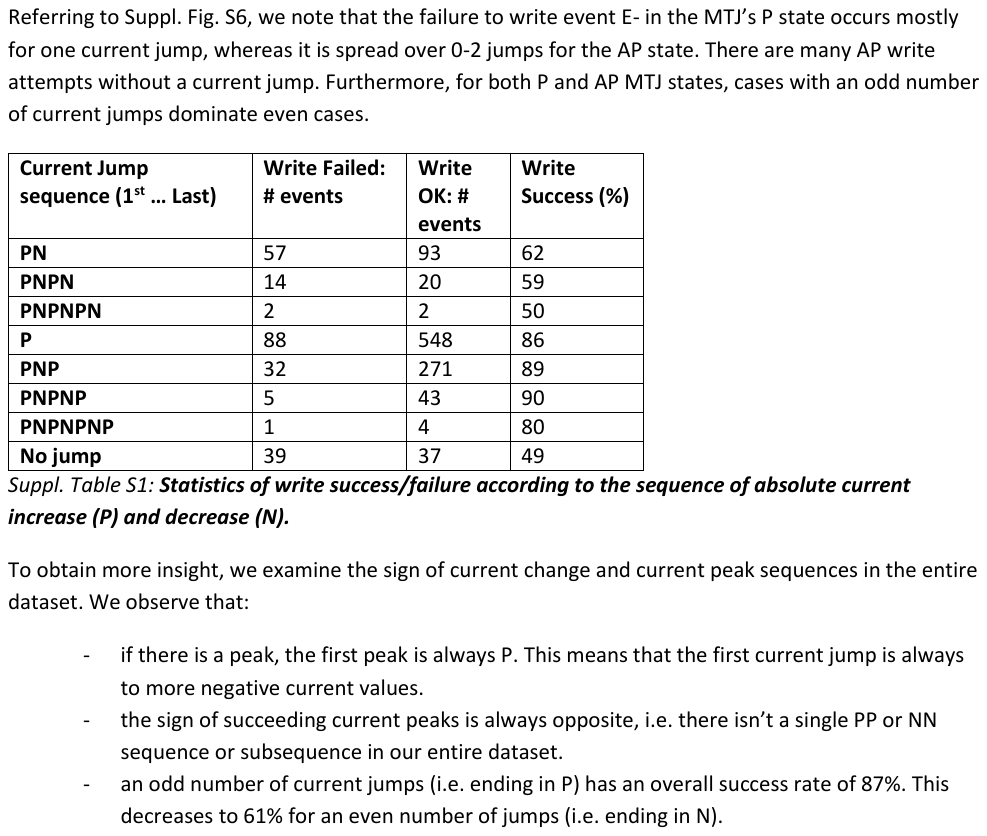}}\end{figure}

\begin{figure}[h]\vspace*{-2cm}\makebox[\textwidth]{\includegraphics[scale=1,page=7]{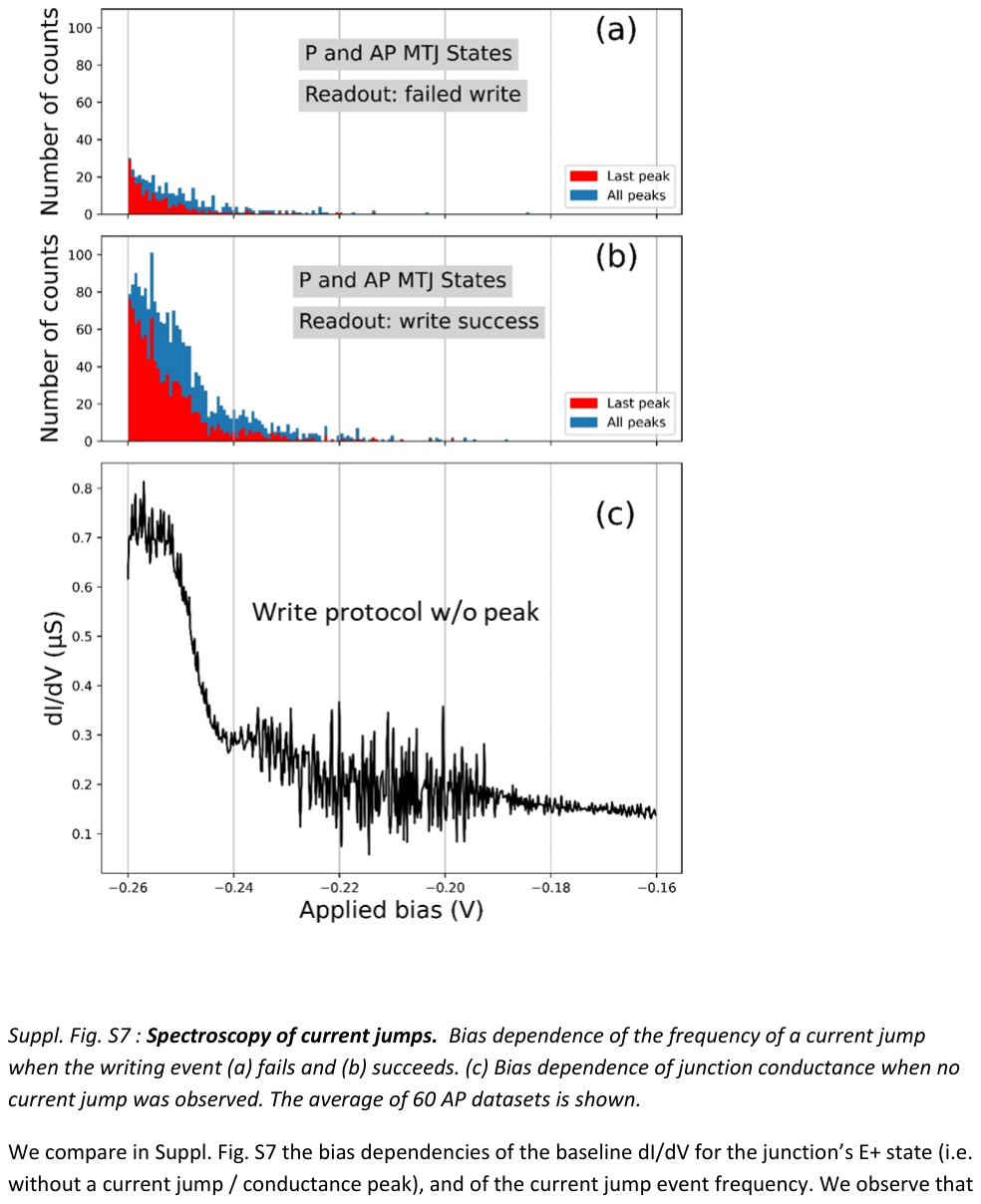}}\end{figure}

\begin{figure}[h]\vspace*{-2cm}\makebox[\textwidth]{\includegraphics[scale=1,page=8]{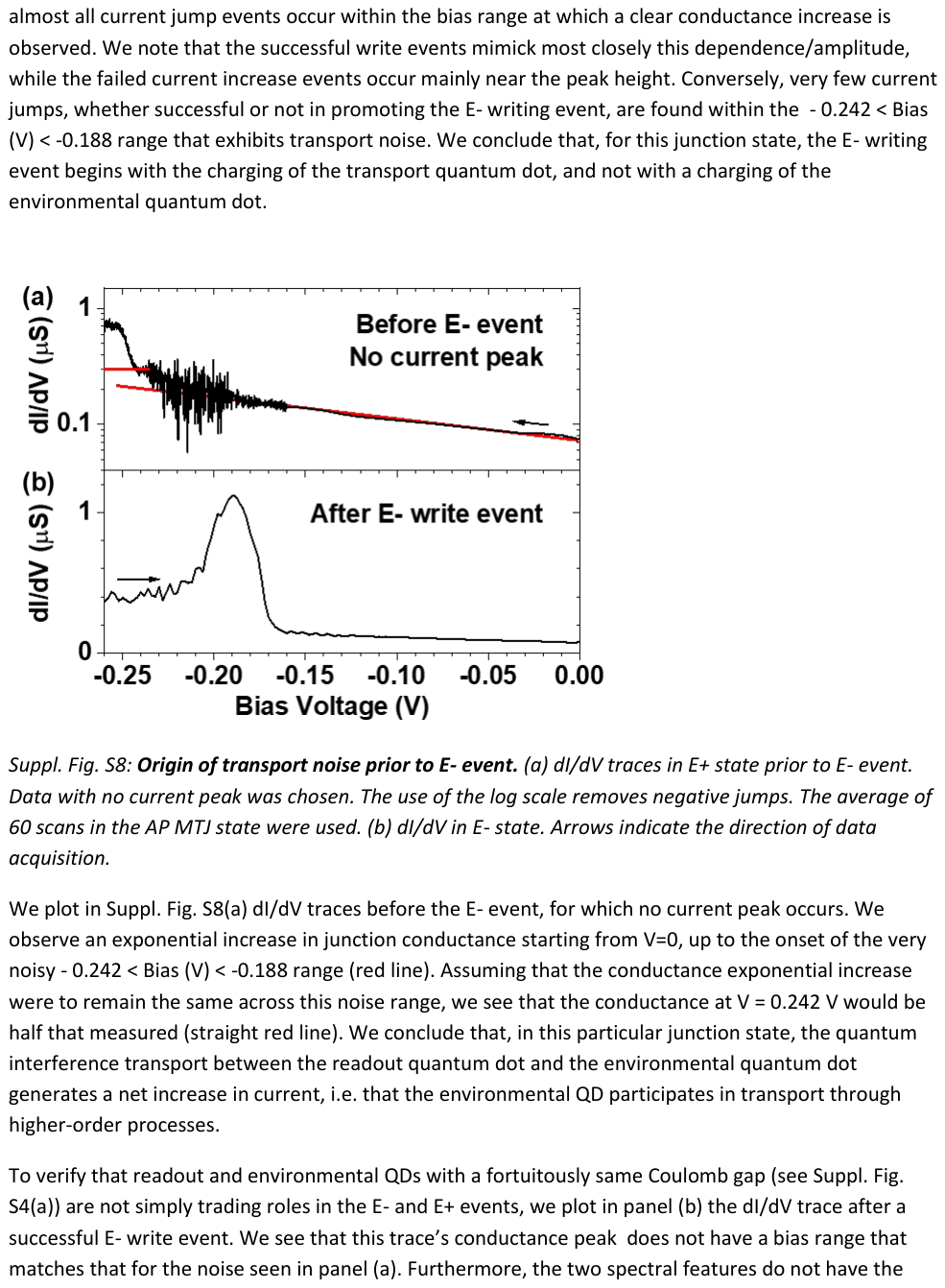}}\end{figure}

\begin{figure}[h]\vspace*{-2cm}\makebox[\textwidth]{\includegraphics[scale=1,page=9]{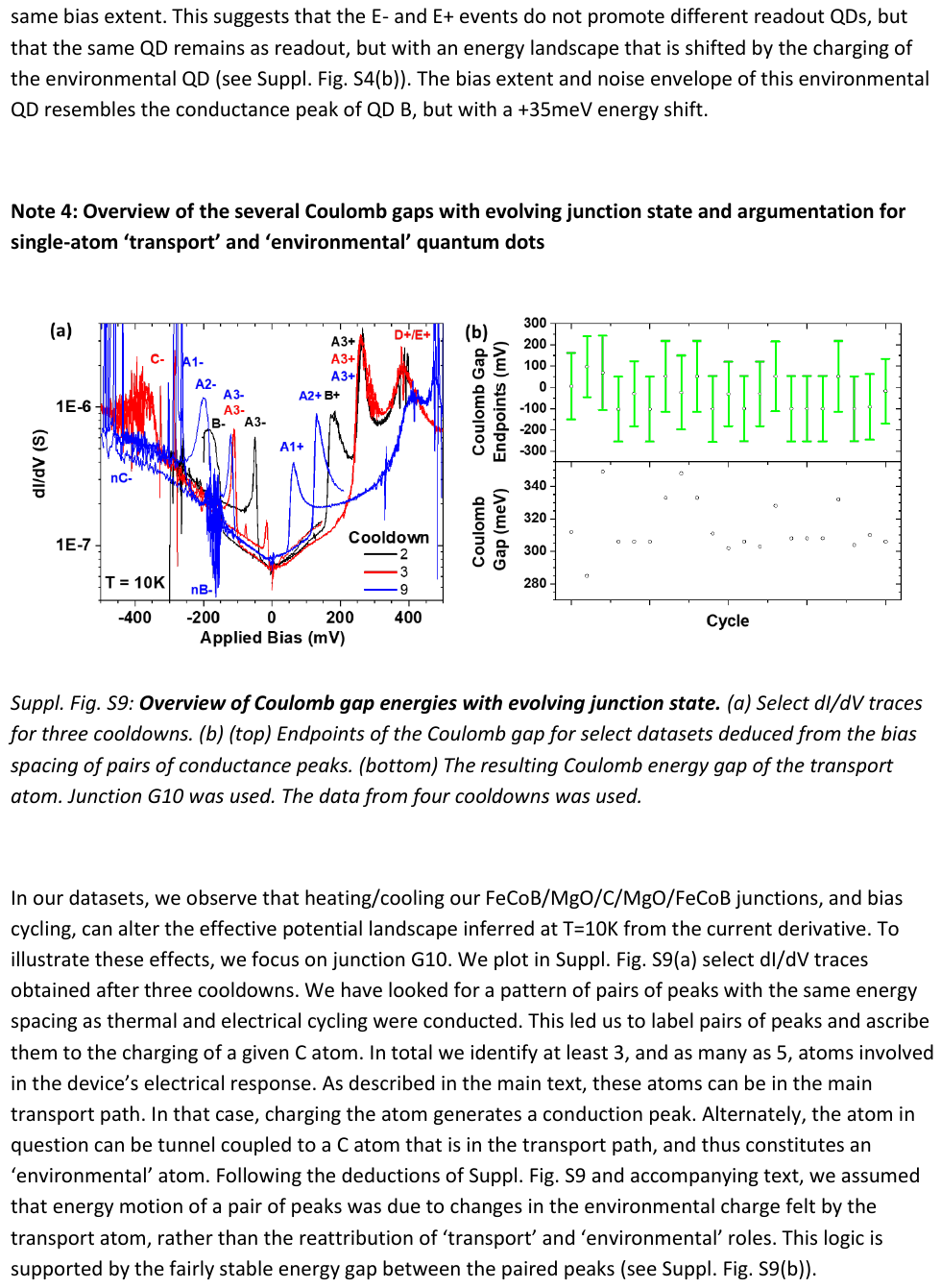}}\end{figure}

\begin{figure}[h]\vspace*{-2cm}\makebox[\textwidth]{\includegraphics[scale=1,page=10]{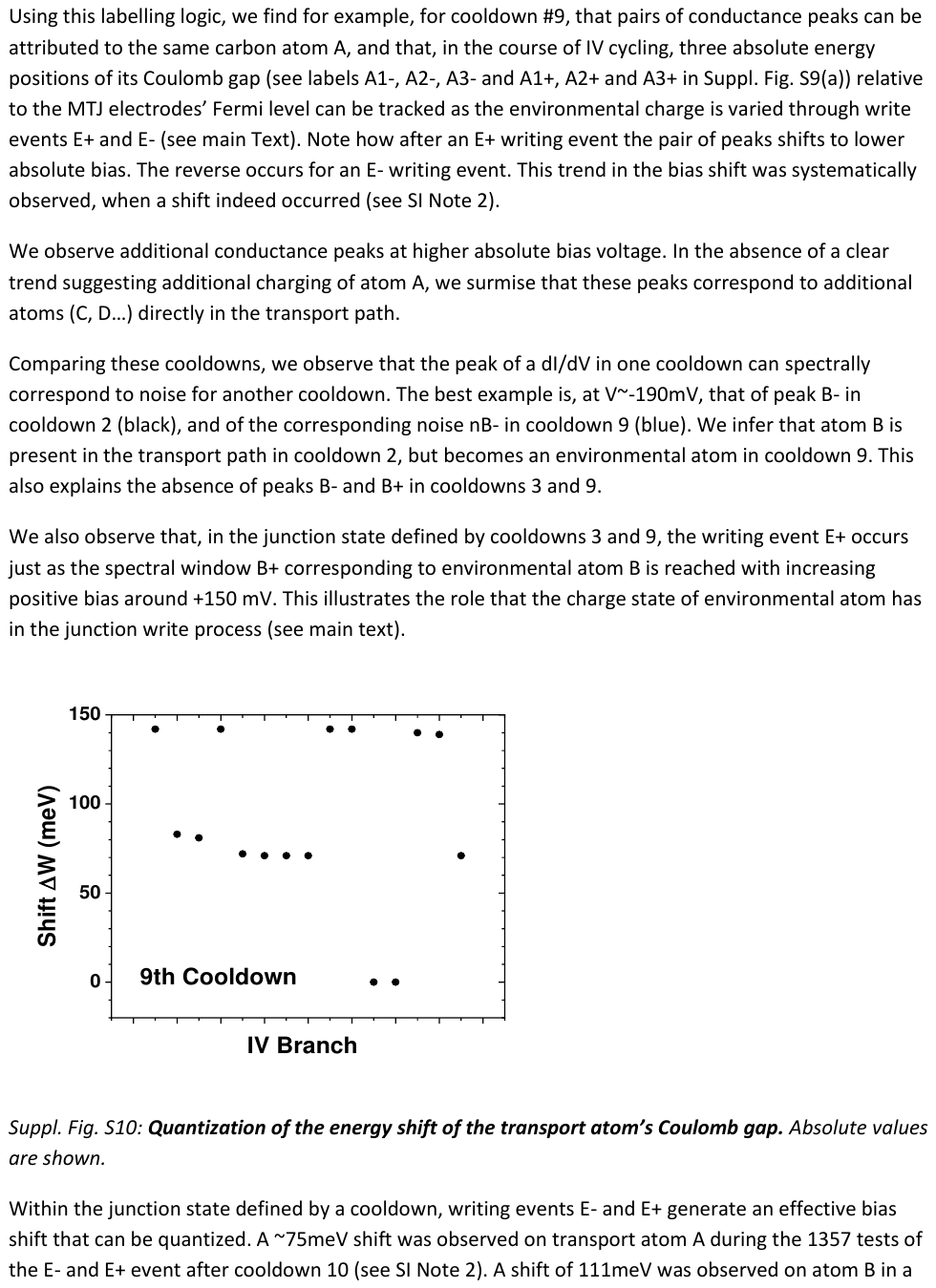}}\end{figure}

\begin{figure}[h]\vspace*{-2cm}\makebox[\textwidth]{\includegraphics[scale=1,page=11]{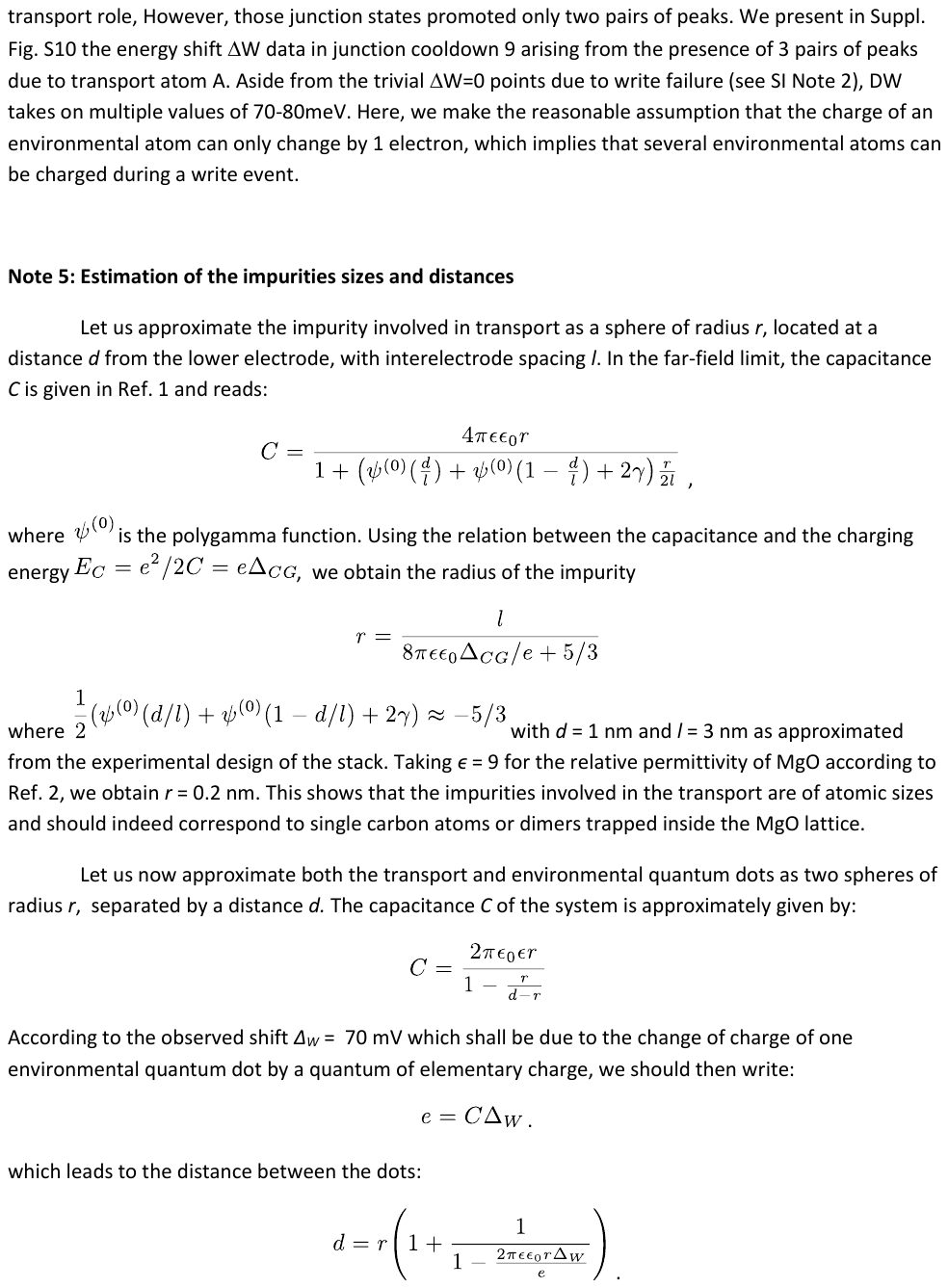}}\end{figure}

\begin{figure}[h]\vspace*{-2cm}\makebox[\textwidth]{\includegraphics[scale=1,page=12]{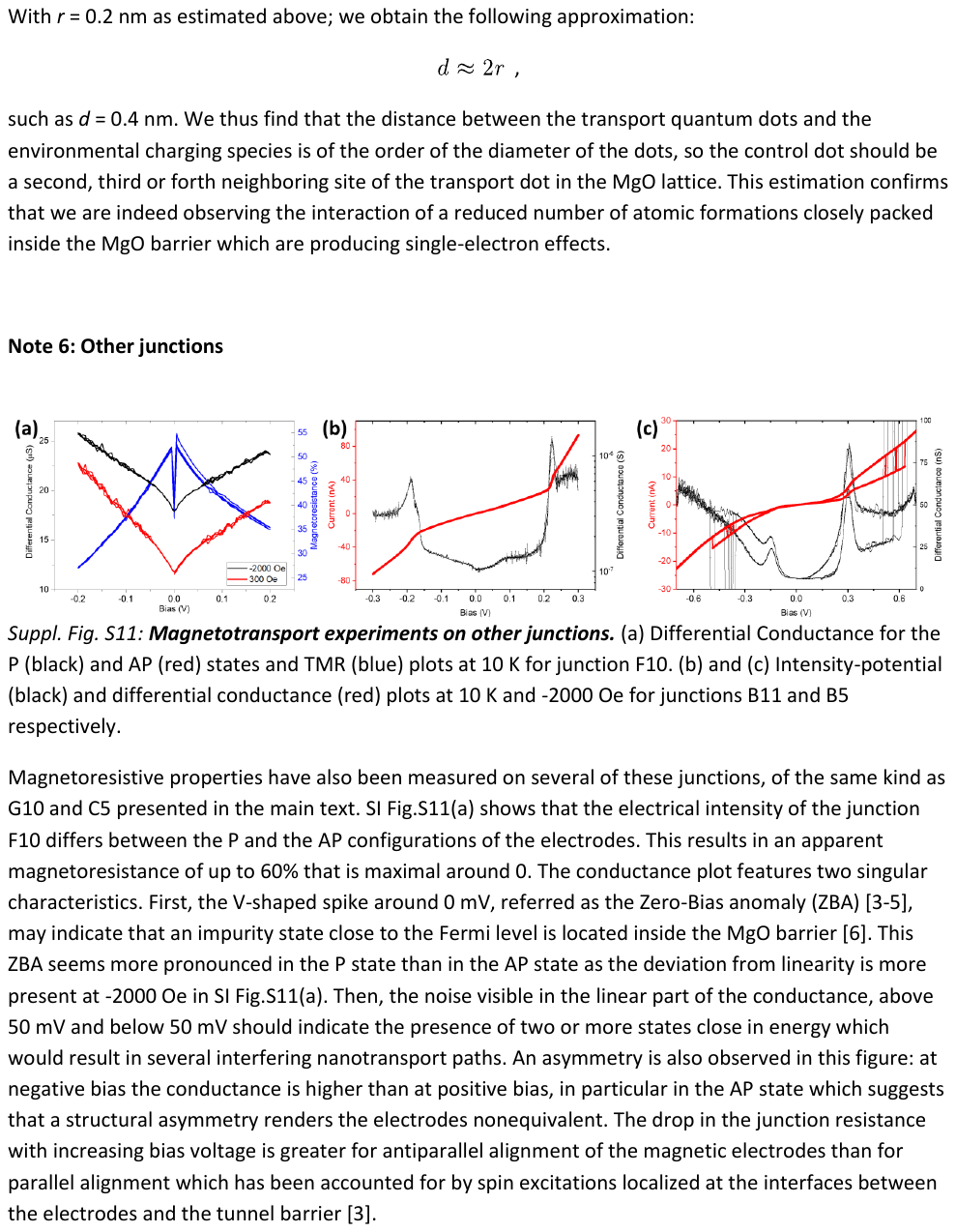}}\end{figure}

\begin{figure}[h]\vspace*{-2cm}\makebox[\textwidth]{\includegraphics[scale=1,page=13]{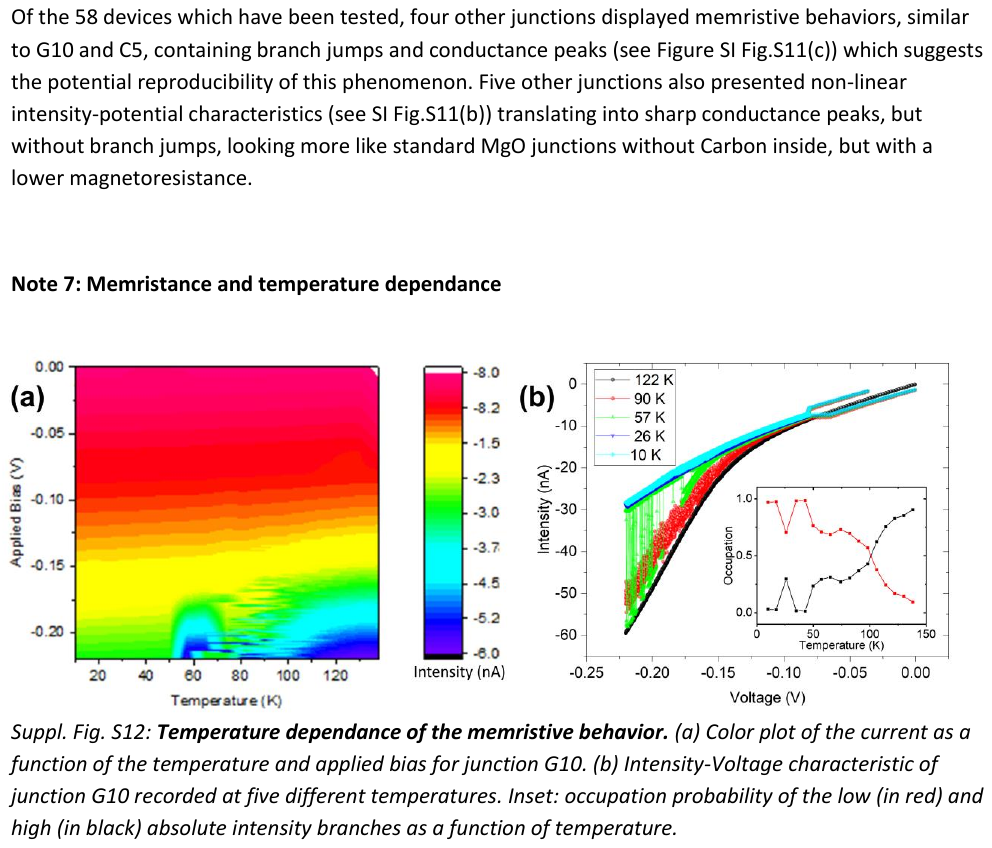}}\end{figure}

\begin{figure}[h]\vspace*{-2cm}\makebox[\textwidth]{\includegraphics[scale=1,page=14]{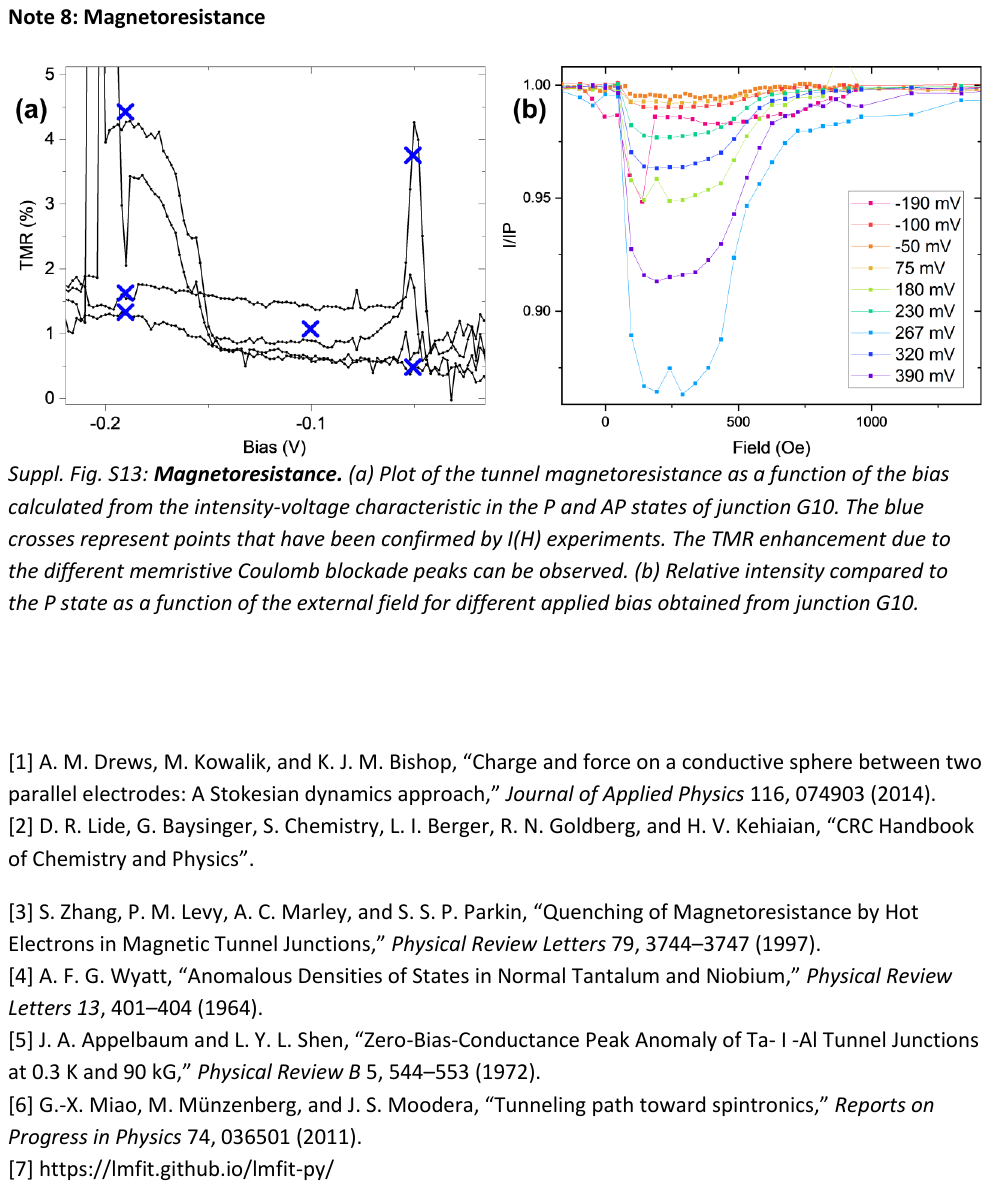}}\end{figure}

\end{document}